\documentclass[apj]{emulateapj}

\shortauthors{Chung et al.}

\begin{document}

\title{Enhanced Nitrogen in Morphologically Disturbed Blue Compact Galaxies at 0.20 $< z <$ 0.35: Probing Galaxy Merging Features}

\author{Jiwon Chung\altaffilmark{1}, Soo-Chang Rey\altaffilmark{1, 4}, Eon-Chang Sung\altaffilmark{2}, Bum-Suk Yeom\altaffilmark{1}, Andrew Humphrey\altaffilmark{3}, Wonhyeong Yi\altaffilmark{1}, Jaemann Kyeong\altaffilmark{2}}
\altaffiltext{1}{Department of Astronomy and Space Science, Chungnam National University, Daejeon 305-764, Korea ; jiwon@cnu.ac.kr, screy@cnu.ac.kr}
\altaffiltext{2}{Korea Astronomy and Space Science Institute,Daejeon 305-348, Korea}
\altaffiltext{3}{Centro de Astrof\'{i}sica da Universidade do Porto, Rua das Estrelas, 4150-762, Porto, Portugal}
\altaffiltext{4}{Author to whom any correspondence should be addressed}
\begin{abstract}
 
We present a study of correlations between the elemental abundances and galaxy morphologies of 91 blue compact galaxies (BCGs) at z=0.20-0.35 with Sloan Digital Sky Survey (SDSS) DR7 data. We classify the morphologies of the galaxies as either `disturbed' or `undisturbed', by visual inspection of the SDSS images, and using the Gini coefficient and $M_{20}$. We derive oxygen and nitrogen abundances using the $T_{e}$ method. We find that a substantial fraction of BCGs with disturbed morphologies, indicative of merger remnants, show relatively high N/O and low O/H abundance ratios. The majority of the disturbed BCGs exhibit higher N/O values at a given O/H value compared to the morphologically undisturbed galaxies, implying more efficient nitrogen enrichment in disturbed BCGs. We detect Wolf-Rayet (WR) features in only a handful of the disturbed BCGs, which appears to contradict the idea that WR stars are responsible for high nitrogen abundance. Combining these results with Galaxy Evolution Explorer (GALEX) GR6 ultraviolet (UV) data, we find that the majority of the disturbed BCGs show systematically lower values of the H$\alpha$  to near-UV star formation rate ratio. The equivalent width of the H$\beta$ emission line is also systematically lower in the disturbed BCGs. Based on these results, we infer that disturbed BCGs have undergone star formation over relatively longer time scales, resulting in a more continuous enrichment of nitrogen. We suggest that this correlation between morphology and chemical abundances in BCGs is due to a difference in their recent star formation histories.
\end{abstract}


\section{Introduction}

Blue compact galaxies (BCGs) are compact, star-bursting galaxies which are thought to be the least chemically evolved ones in the local Universe, and are therefore considered as analogs of high-redshift star-forming galaxies. They allow important constraints to be placed on massive star formation (SF) and the evolution of galaxies in very low metallicity environments (Thuan \& Martin 1981; Izotov \& Thuan. 1999). While a number of studies have probed the formation and evolution of BCGs, mergers and interactions between various types of galaxy have been suggested as mechanisms to stimulate enhancement of SF activity in BCGs (e.g., Larson \& Tinsley 1978; Taylor et al. 1995, Mendez \& Esteban 2000; Pustilnik et al. 2001a, b; {\"O}stlin et al. 2001; Sung et al. 2002; Johnson et al. 2004; Ekta \& Chengalur 2010a). Recently, based on numerical simulations, Bekki (2008) showed that blue compact dwarf (BCD) galaxies  can be formed by mergers and interactions between gas-rich dwarfs. In his simulations, mergers are able to trigger a central starburst to form young, massive cores, resembling the morphologies of currently known BCD galaxies. 

The gas phase elemental abundances of BCGs is a key physical parameter which traces their star formation history (SFH) and chemical evolution. The N/O abundance ratio as a function of the O/H ratio is considered to be an important observational constraint for probing the chemical evolution of galaxies (Edmunds \& Pagel 1978; Torres-Peimbert, Peimbert, \& Fierro 1989; Vila-Costas \& Edmunds 1993; Izotov et al. 2006). There has been observational evidence that the oxygen and nitrogen abundances may be profoundly linked to the local environments of galaxies. Many studies have suggested that mergers and tidal interactions can induce a low O/H ratio in the galaxies' central region owing to inflows of metal-poor gas from the outer part of the galaxy (Rupke et al. 2008, 2010; Peeples et al. 2009; Ekta \& Chengalur 2010b; Kewley et al. 2010; Montuori et al. 2010). While the N/O ratio was found to be fairly constant for low-metallicity BCGs (Izotov \& Thuan 1999), a nitrogen overabundance was detected in some BCGs. Furthermore, some studies suggested that the unexpectedly high N/O ratios of BCG could be associated with the merger process and hence with the related powerful starburst (e.g., Pustilnik et al. 2004). Recently, based on a sample of star-forming luminous compact galaxies at z $<$ 0.63, Izotov, Guseva, \& Thuan (2011) found a significant spread in N/O values, with a hint of enhanced nitrogen abundance. Furthermore, in their sample, a considerable fraction of the galaxies show signs of having disturbed morphologies, suggesting recent interactions.

All of these previous results suggest that the effects of mergers and interactions provide a remarkable opportunity for clarifying the origin of the scatter in the N/O vs. O/H diagram. However, despite the importance of elemental abundances, detailed studies of their relationship with the morphology and the SFH of BCGs have been limited by small samples. To improve this situation, in this letter, we examine a large and homogeneous sample of BCGs at intermediate redshifts around z$\sim$0.3. Our goal is to assess whether the BCGs are discriminated by features of their elemental abundances, linking their SF processes to the merger and interaction of galaxies.

\section{Sample Selection and Analysis}

We select a sample of 91 highly excited BCGs in the redshift range of 0.20 $<$ z $<$ 0.35 using criteria that are based on both spectra and images from the Sloan Digital Sky Survey (SDSS) Data Release 7 (DR7) (Abazajian et al. 2009). Only galaxies with a well-detected [OIII]${\lambda}$4363 emission line, with S/N higher than 5, are selected. This criterion allows a direct determination of the electron temperature and abundances. We select only galaxies which occupy the region of star-forming galaxies with the highest excitation and which are well separated from active galactic nuclei in the BPT diagram (Baldwin, Phillips, \& Terlevich 1981).
In addition, only BCG galaxies with relatively compact morphologies and blue colors in their SDSS images are considered. Our BCGs span a luminosity range of -22 $<$ $M_{r}$ $<$ -19 and color ranges of 0.1 $<g-r<$ 1.2 and -2.0 $<r-i<$ 0.2. According to the photometric selection criteria used by Cardamone et al. (2009), about 59 \% (54 of 91) of our BCGs are classified as green pea galaxies. Our sample also includes 14 BCGs which show Wolf-Rayet (WR) features based on the detection of a blue bump around 4650${\AA}$ and/or the He II emission line at 4686${\AA}$ (Conti 1991; Pustilnik et al. 2004; Brinchmann, Kunth, \& Durret 2008). 

By visual inspection of the SDSS $g, r, i$ combined color images, we conducted a systematic morphological analysis in which we classified each galaxy as either disturbed or undisturbed. We classified 19 galaxies as disturbed, on account of their asymmetric and distorted outer light distribution and/or cometary appearance, indicative of merger remnants or recent disturbance by tidal interaction. The 72 BCGs classified as undisturbed show compact morphologies of relatively round and smooth shape. Example SDSS images for undisturbed and disturbed BCGs are shown in Figure 1. For the four undisturbed BCGs in our sample which also have an optical HST or SDSS Stripe 82 image, we have tested the impact of image quality on our morphological classification (see inset images of Fig. 1a, b for two objects). None of these higher-quality, ancillary images show clear evidence of mergers or interactions, although one does show an elongated shape  with a hint of two separate SF regions (see Fig. 1a). Note that the disturbed BCGs are systematically biased towards higher luminosities (-22 $<$ $M{_r}$ $<$  -21), while undisturbed BCGs span the whole luminosity range of our sample.

In addition to our visual classification, we have derived the Gini coefficient and $M_{20}$ from the SDSS r-band images, using the method of Lotz, Primack, \& Madau (2004).  In G-$M_{20}$ parameter space (Figure 2), the disturbed and undisturbed BCGs show clearly different distributions.  In particular, the disturbed BCGs show higher G and lower $M_{20}$.  In other words, the disturbed BCGs show steeper, and thus more centrally-concentrated intensity distributions than the undisturbed BCGs, which is the result of distorted low surface-brightness regions in the outskirts of the disturbed galaxies. This result supports our visual morphological classification.

From the Galaxy Evolution Explorer (GALEX) GR6 data, 35 BCGs are detected in the far-ultraviolet (FUV; 1350-1750${\AA}$) and the near-ultraviolet (NUV; 1750-2800${\AA}$) bands. The FUV and NUV fluxes are taken from the GALEX pipeline data. Based on the reddening law of Cardelli et al. (1989), galactic extinction was corrected using the E(B-V) value from the NED. The internal reddening coefficient C(H$\beta$) was derived from the SDSS spectrum and its absorption was determined from A(FUV) = 8.19 ${\times}$ E(B-V) = 5.72/C(H$\beta$) and A(NUV) = 9.21 ${\times}$ E(B-V) = 6.44/C(H$\beta$).

The global fluxes of the emission lines from each galaxy were systematically measured by fitting a Gaussian profile, using the IRAF $splot$ task. We applied a foreground reddening correction to the line intensities following Lee \& Skillman (2004). Using the traditional Balmer decrement method (Izotov et al. 1994), the observed line intensities were also corrected for both internal reddening and underlying stellar absorption. The gas-phase elemental abundances of O and N were derived by the direct method, using the STSDAS $nebular/ionic$ task. Using the ionization correction factors from Izotov et al. (2006), we corrected for unobserved ionization stages and then estimated the total elemental abundances. 

We derive the {H$\alpha$}, FUV, and NUV star formation rates (SFRs) of our BCGs using the relations given by Kennicutt et al. (1998). We have corrected the H$\alpha$ flux to take into account light-loss from the 3 arcsec diameter of the SDSS spectroscopic aperture.  The total H$\alpha$ flux was estimated using the ratio between Petrosian and 3 arcsec aperture fluxes in the i-band surface photometry. The SDSS aperture collects on average 80\% of the total H$\alpha$ flux of our galaxies, owing to the compactness of the galaxies (i-band Petrosian radii are in the range 0.9" - 3.3"). In Table 1, we summarize the median values of basic parameters derived from our analysis, for disturbed and undisturbed BCGs.

In addition, a sample of 104 local BCGs, with a similar luminosity range to our BCG sample, was compiled from the literature (Izotov \& Thuan 1999; Izotov et al. 2006) in order to compare the properties of BCGs at low and intermediate redshifts. Using SDSS DR7 images, we classified the morphologies of the local BCGs into disturbed and undisturbed galaxies as described above. The gas-phase abundances of local BCGs are also adopted from Izotov \& Thuan (1999) and Izotov et al. (2006). For comparison with our disturbed BCGs, we also selected five well-known examples of local BCGs showing clear merger features (HS 0837+4717, III Zw 107, UM420, SBS 0926+606, and IRAS 08208+2816; Pustilnilk et al. 2004; Lopez-Sanchez 2010). All of these local merger BCGs are detected by GALEX, from which we derive their FUV and NUV SFR. The H$\alpha$ SFR of all but one (HS 0837+4717) are available from Kennicutt et al. (2008). 

\section{Results}

\subsection{N/O and O/H ratios}
In Figure 3, we show N/O vs. O/H for the disturbed (red filled circles) and undisturbed (black filled circles) BCGs. For comparison, we also show the distributions of disturbed (solid contours) and undisturbed (dashed contours) local BCGs. The BCGs show metallicities in the range of 7.6 $<$ 12+log O/H $<$ 8.5 with a median value of 8.13 $\pm$ 0.37. This corresponds to  $\sim$ 27\% of the solar value assuming 12+log(O/H)$_\odot$ = 8.69 (Allende Prieto et al. 2001), indicating that these systems are genuine metal-poor galaxies. In the case of the galaxies with 12+log(O/H) $>$ 8.0, the overall distribution of our BCGs shows no prominent difference to that of local BCGs, indicating no segregation of elemental abundances of BCGs at z $<$ 0.35 (see also Izotov et al. 2011). On the other hand, it is worth noting that our BCGs with low O/H values (12 + log O/H $<$ 8.0) appear to have systematically higher N/O values compared to those showing relatively higher O/H values. 

Regarding the distribution of elemental abundances as a function of morphology, one remarkable result is that the disturbed BCGs exhibit a clearly different distribution to that of the undisturbed ones. Most of the disturbed BCGs display systematically higher N/O values at a given O/H value compared to the undisturbed ones. The mean difference of N/O values between two subsamples is very pronounced (about 0.2 dex, see Table 1 and histograms in the right panel of Fig. 3). A K-S test on the N/O distributions between disturbed and undisturbed BCGs yields a probability of 0.04 \% for the same parent distribution of the two subsamples. At the same time, a large fraction of the disturbed BCGs is also systematically biased towards low O/H values (12+log(O/H) $<$ 8.0). In line with previous results (Rupke et al. 2008, 2010; Peeples et al. 2009; Ekta \& Chengalur 2010b; Kewley et al. 2010; Montuori et al. 2010), the low O/H values of some of the disturbed BCGs support the idea that these galaxies are associated with recent mergers or interactions with neighboring galaxies. When we consider only the disturbed BCGs, there appears to be an anticorrelation between the N/O and O/H ratios such that disturbed BCGs with low O/H values show systematically high N/O values. 

Note that the five local disturbed BCGs (blue squares in Fig. 3) exhibit a similar distribution to our disturbed BCGs, and most of them also have high N/O values. The most extreme object is HS 0837+4717, which is a very metal-deficient BCG (12+log(O/H)=7.64) showing a significant N excess (log(N/O)= -0.83: Pustilnilk et al. 2004). 

\subsection{Wolf-Rayet Features}

The bias towards high N/O values in morphologically disturbed BCGs necessitates mechanisms for the enhanced production of nitrogen in these galaxies. One possibility is that WR stars are formed in starbursts triggered by mergers. Since merging of two galaxies provides a subsequent intense starburst, this could result in a significant increase of the relative fraction of massive young stars. Four of the five local merger BCGs have previously been shown to have clear WR features (blue squares with yellow cross in Fig. 3; Pustilnilk et al. 2004; Lopez-Sanchez 2010, but see James, Tsamis, \& Barlow 2010 for non-detection of WR features in UM420 from their IFU observations). However, WR features are detectable for only a handful of our disturbed BCGs (four red filled circles with yellow cross in Fig. 3). Furthermore, WR features are not preferentially found in the sample of disturbed BCGs with high N/O values, regardless of morphology.

On the other hand, three of the four disturbed BCGs with WR features appear to show morphologies more typical of an on-going merger or approaching post-merger (i.e., contact between two separate galaxies or distinct tidal tail, see Fig. 1d, e, f), although it is somewhat difficult to properly assess detailed morphologies with just the SDSS images. In the case of the local BCGs with WR features, they also show similar, on-going merger morphologies (e.g., two separate galaxies for III Zw 107 and UM420 as well as a distinct, long tidal tail for IRAS 08208+2816). This suggests that the appearance of WR features in disturbed BCGs might be related to the degree of disturbance in their morphologies, which itself is an indicator of merger-status and, therefore, starburst activity.

The detection and discrimination of WR features in the spectrum of a star burst galaxy can be hampered by several effects, including the quality of the spectrum, or the location and size of the spectroscopic aperture (see Lopez-Sanchez \& Esteban 2008 and references therein). In the case of large apertures, weak WR features can be strongly diluted by the continuum flux (Lopez-Sanchez \& Esteban 2008; James et al. 2010). For this reason, we cannot rule out the existence of WR populations in the disturbed BCGs, even though they are not clearly detected in the SDSS spectra.

\subsection{Star Formation Histories}
The starburst duration directly affects the chemical yield and the stellar age of a galaxy. Long bursts of SF are able to produce high N/O values, without necessarily showing distinct WR features. While WR stars are widely separated in time, resulting in a weak or no WR features in the spectrum, the interstellar medium (ISM) in BCGs is cumulatively enriched over time by massive stars. Since the disturbed BCGs in our sample are systematically biased towards higher masses (-22 $<$ $M{_r}$ $<$ -21) (see also Telles, Melnick, \& Terlevich, 1997), they can retain a large fraction of newly produced metals. The relatively long duration of the SF for BCGs that are formed by the merger of very gas-rich dwarfs is supported by numerical simulations (200 - 300 Myr, see Fig. 4 of Bekki 2008). 

Different indicators of the SFR are sensitive to stars of different masses and timescales.  The H$\alpha$ emission traces the present SF activity from massive, early type stars with timescales of a few tens of Myr. On the other hand, the UV flux traces less massive stars with timescales of a few hundred Myr (Kennicutt 1998; Iglesias-Paramo et al. 2004). The H$\alpha$ luminosity declines more rapidly than the UV continuum luminosity, which leads to a decrease in the H$\alpha$ to UV SFR ratio with increasing starburst age (Sanchez-Gil et al. 2011). Thus, the H$\alpha$ to UV SFR ratio carries important information about the SFH and the duration of the SF episode of a galaxy.  In Figure 4 (left panels), there is a striking difference between the disturbed (red filled circles) and undisturbed BCGs (black filled circles), in terms of their H$\alpha$ to NUV SFR ratios; the disturbed BCGs show SFR ratios that are about a factor of 5 lower than the undisturbed BCGs (see also Table 1). Interestingly, the nearby BCGs that show merger-related features (blue squares) are also consistent with this result. This supports the idea that disturbed BCGs have undergone SF over relatively longer time scales than the undisturbed BCGs. 

The H$\beta$ emission line equivalent width, EW(H$\beta$), has also been suggested as an indicator of the age of the latest burst of SF in a galaxy (Stasinska \& Izotov 2003; Izotov et al. 2011). It is clearly seen that the majority of the disturbed BCGs also show systematically lower EW(H$\beta$) than the undisturbed ones (about a factor of 3), implying a significant underlying, non-ionizing older stellar population which contributes to the stellar continuum (see right panels of Fig. 4 and Table 1). This result provides further support for the idea that disturbed and undisturbed BCGs have different SFH.

\section{Discussion and Conclusions}

We have established that the chemical abundances within BCGs at 0.20 $<$ z $<$ 0.35 are closely correlated with galaxy morphology. The BCGs in our sample are segregated by their morphology in the N/O vs. O/H diagram. While there have been several observational and theoretical suggestions that galaxy mergers or interactions result in a decrease in O/H, our study has shown that a significant fraction of morphologically disturbed BCGs also exhibits nitrogen enhancement. Disturbed BCGs with an elevated nitrogen abundance might have experienced prolonged SFHs through dynamical events (e.g., merger or interaction) between two galaxies, and consequently the ISM is continuously enriched by massive stars. Considering the high fraction of green pea galaxies in our BCGs, our results are consistent with recent studies of abundance patterns in green pea galaxies, which have revealed systematically higher N/O ratios compared to most star forming galaxies at a given O abundance, possibly resulting from recent interaction-induced inflows of gas (Amorin et al. 2010, 2012). Inflow of very low metallicity gas may dilute O/H, without significantly affecting the N/O ratio. This might explain the simultaneously high N/O and relatively low O/H ratios of some of the BCGs, due to the horizontal offset towards lower O/H values in the N/O vs. O/H diagram (K{\"o}ppen \& Hensler 2005). On the other hand, it is important to point out that our results are not affected by the difference in luminosity of the disturbed and undisturbed subsamples. If we instead consider only the BCGs in the luminosity range -22 $<$ $M{_r}$ $<$ -21, the results of our analysis are not significantly changed.

Our results pertain to the presence of the strongest optical morphological signatures of merging or interaction. However, due to the limited spatial resolution (i.e., 8 kpc resolution for the 1.4 arcsec SDSS PSF at z $\sim$ 0.3) and depth of the SDSS images used, we cannot rule out that we have missed faint merging/interaction morphological features in some of the apparently undisturbed BCGs. Based on the trends identified in our sample, objects with low level, small scale merging or interaction features may appear as morphologically undisturbed in the SDSS images but with a high N/O, low H$\alpha$ to UV SFR ratio, and low EW(H$\beta$). Further deep, high resolution imaging observations will be necessary for clarifying this issue.

\acknowledgments
We thank the anonymous referee for thoughtful comments that helped to improve this paper.
This research was supported by Basic Science Research
Program through the National Research Foundation of
Korea (NRF) funded by the Ministry of Education, Science and
Technology (2012R1A1B4003097). Support for this work was
also provided by the NRF of Korea to the Center for Galaxy
Evolution Research. AH acknowledges a Marie Curie Fellowship.

\clearpage

\begin{figure}
\epsscale{1}
\plotone{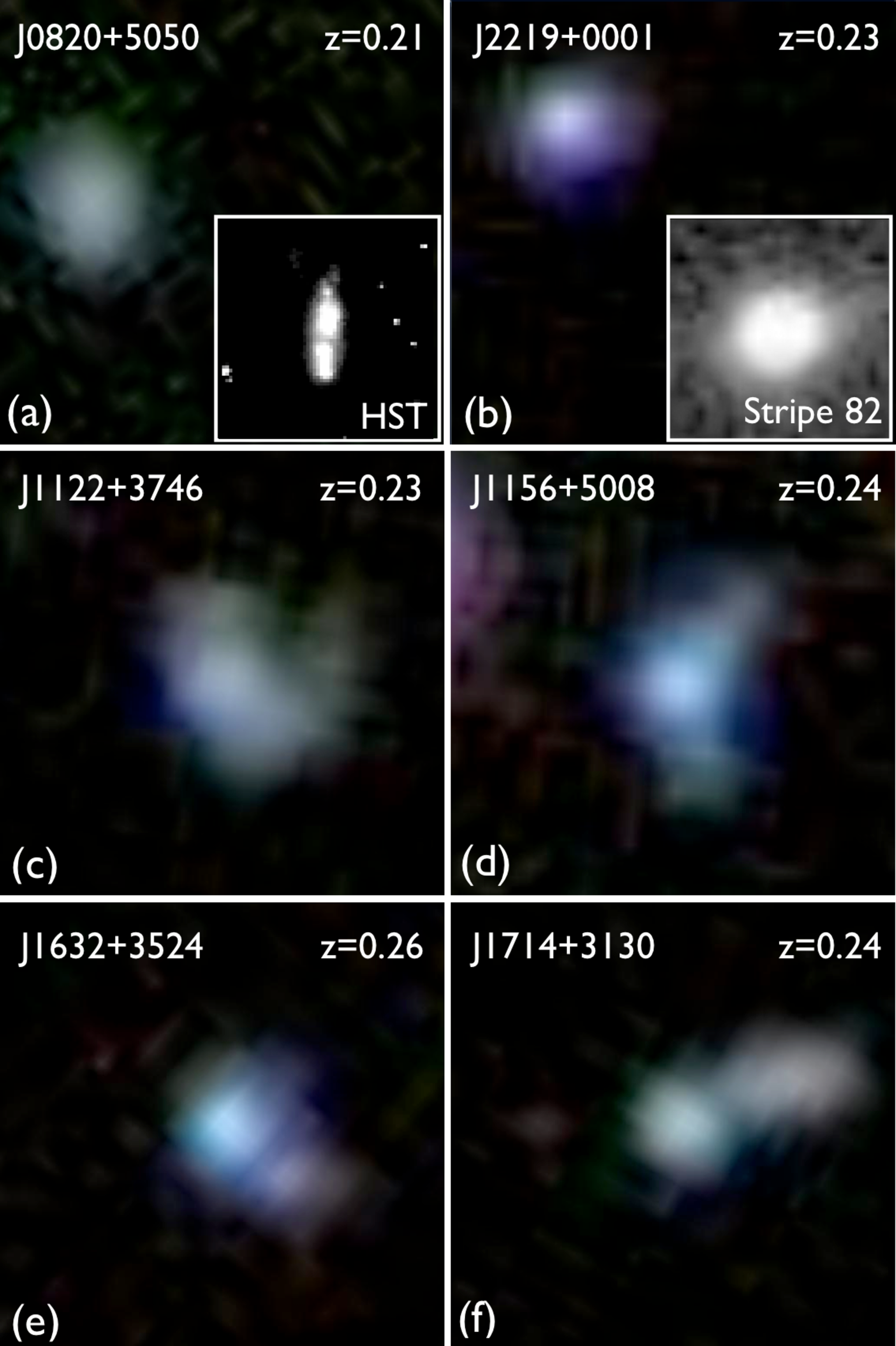}
\begin{center}
\caption {Example of 12.6$''$  ${\times}$ 12.6$''$ SDSS $g, r, i$ composite color images of morphologically undisturbed and disturbed BCGs: (a, b) undisturbed BCGs with round and compact shapes. Inset images are those from HST/WFPC2 or SDSS Stripe 82 field; (c) disturbed BCG without WR features in its spectrum; (d, e, f) disturbed BCGs with WR features in their spectra. In each panel, the SDSS name and redshift of the galaxy is indicated.\label{fig1}}
\end{center}
\end{figure}
\clearpage

\begin{figure*}
\epsscale{1}
\plotone{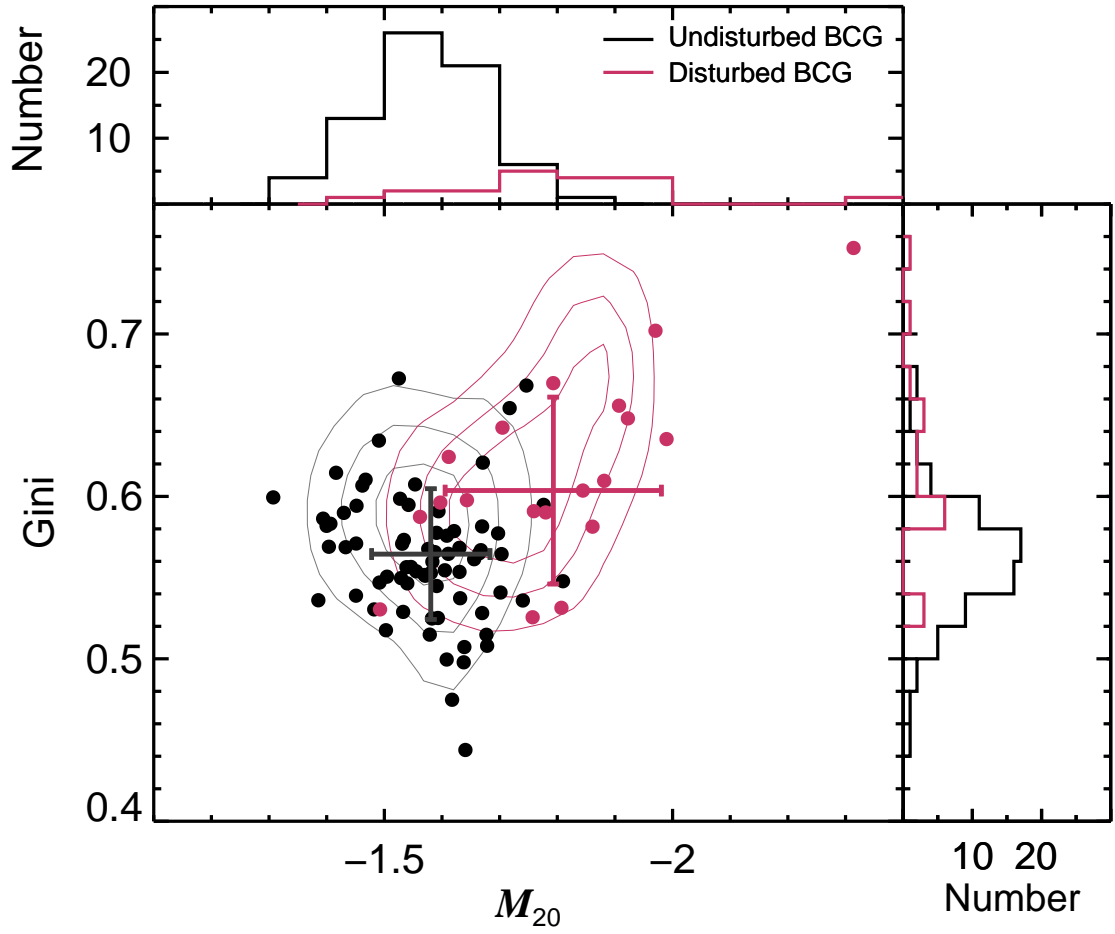}
\begin{center}
\
\
\
\caption {G vs. $M_{20}$ diagram for disturbed (red circles and contours) and undisturbed (black circles and contours) BCGs. Error bars indicate the mean values and standard deviations of G and $M_{20}$ for these two subsamples, while the histograms show their distributions. The mean values of G and $M_{20}$ are [0.60, -1.79] and [0.56, -1.58] for disturbed and undisturbed BCGs, respectively. \label{fig2}}
\end{center}
\end{figure*}
\clearpage

\begin{figure*}
\epsscale{1}

\plotone{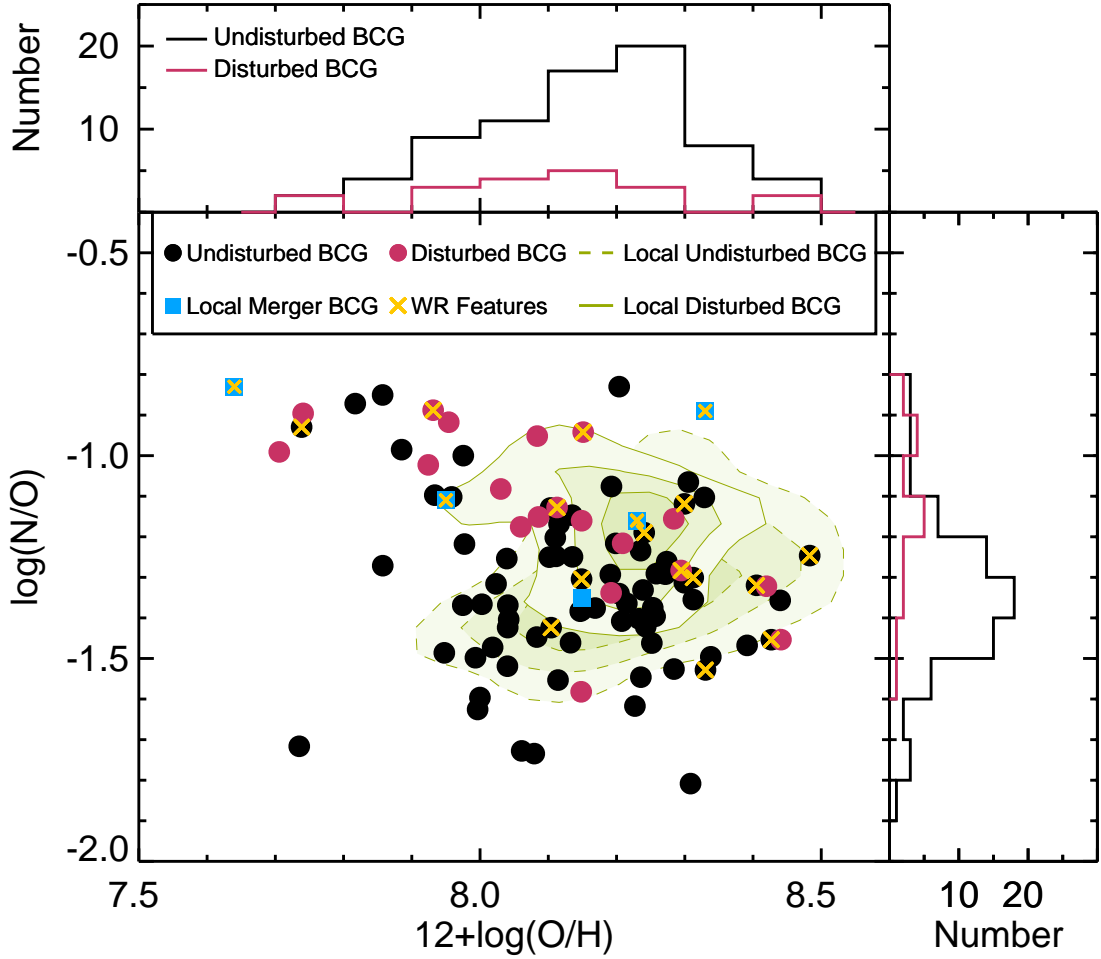}
\begin{center}
\
\caption{N/O vs. O/H diagram of BCGs. Red and black filled circles are disturbed and undisturbed BCGs, respectively. The distributions of the disturbed and undisturbed local BCGs from the literature (Izotov \& Thuan 1999; Izotov et al. 2006) are denoted by solid and dashed contours, respectively. Blue squares represent the local merging BCGs. BCGs with WR features are represented by crosses. The histograms show the distributions of N/O and O/H for the disturbed and the undisturbed BCGs.\label{fig3}}
\end{center}
\end{figure*}
\clearpage

\begin{figure*}
\epsscale{1}
\begin{center}
\plotone{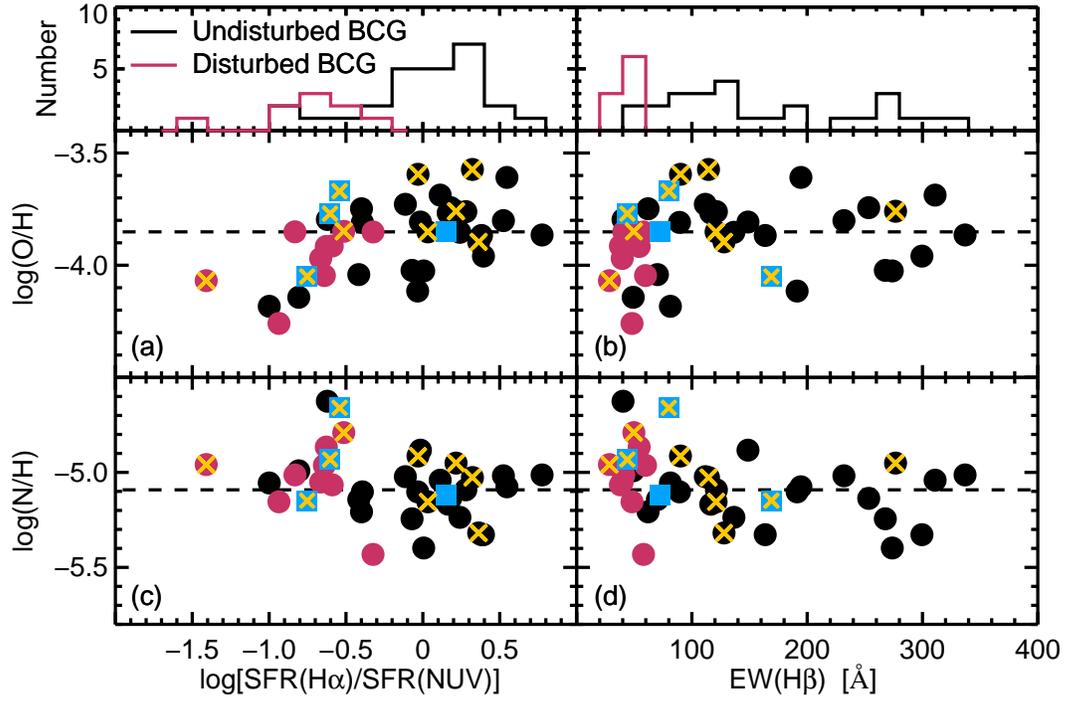}

\caption{(a, c) Elemental abundance vs. the H$\alpha$ to NUV SFR ratio of the BCGs. (b, d) Elemental abundance vs. H$\beta$ emission line equivalent width, EW(H$\beta$), of the BCGs. The dashed line in each panel presents the median value of the undisturbed BCGs. Symbols are the same as in Figure 3. The histograms show the distributions of H$\alpha$ to NUV SFR ratio and EW(H$\beta$) for the disturbed and undisturbed BCGs.}\label{fig4}
\end{center}
\end{figure*}
\clearpage

\begin{center}
  \begin{deluxetable}{ccccccccc} 
  \small\addtolength{\tabcolsep}{-2pt}
 \tabletypesize{\tiny}
  \tabletypesize{\small}
  \tabletypesize{\footnotesize}
 \tabletypesize{\scriptsize}
  \tablenum{1}
  \tablecolumns{10}
  \tablewidth{0pc}
  \tablecaption{Average characteristics of the disturbed and undisturbed BGCs \label{tab1}} 

   \tablehead{
 \colhead{Sample}  & \colhead{12+log(O/H)}  &  \colhead{log(N/O)}& \colhead{SFR$_{H\alpha}$} &  \colhead{SFR$_{NUV}$} & \colhead{log [SFR$_{H\alpha}$/SFR$_{NUV}]$} & \colhead{EW(H$\beta$)}   \\
\colhead{} & \colhead{ }  & \colhead{ } & \colhead{[M$_\sun$/yr]} & \colhead{[M$_\sun$/yr]}& \colhead{ }   & \colhead{[$\AA$]}  & }

\startdata

Disturbed BCG & 8.11 & -1.15 & 10.91 & 50.43 & -0.64 & 48  &   \\  
Undisturbed BCG & 8.15 & -1.35 & 6.99 & 8.82 & 0.11 & 136 & \\ 
 \enddata

  \end{deluxetable}
\end{center}

\end{document}